\documentclass[twocolumn,floats,amsmath,amssymb,pre]{revtex4}

\usepackage{graphicx}

\begin{document}

\title{Path-integral simulation of ice VII: Pressure and temperature effects}
\author{Carlos P. Herrero}
\author{Rafael Ram\'irez}
\affiliation{Instituto de Ciencia de Materiales de Madrid,
         Consejo Superior de Investigaciones Cient\'ificas (CSIC),
         Campus de Cantoblanco, 28049 Madrid, Spain }
\date{\today}

\begin{abstract}
The effects of pressure and temperature on structural and thermodynamic 
properties of ice VII have been studied by using path-integral molecular 
dynamics (PIMD) simulations.  
Temperatures between 25 and 450 K, as well as pressures 
up to 12 GPa were considered. 
Interatomic interactions were modeled by using the effective q-TIP4P/F 
potential for flexible water.
We analyze the pressure dependence of the molar volume, bulk
modulus, interatomic distances, kinetic energy, and atomic delocalization
at various temperatures.
Results of PIMD simulations are compared with those derived from a
quasi-harmonic approximation (QHA) of vibrational modes, which helps to 
assess the importance of anharmonic effects, as well as the influence of 
the different modes on the properties of ice VII.
The accuracy of the QHA for describing this high-pressure phase decreases
for rising temperature, but this approximation becomes more reliable as
pressure grows, since anharmonicity becomes less relevant.
Comparisons with low-pressure cubic ice are presented. \\

\end{abstract}

\maketitle

\section{Introduction}

Over the last few decades the phase diagram of water in the high-pressure 
region has attracted great interest in condensed matter physics and 
chemistry, as well as in planetary sciences \cite{ei69,pe99,fr00,ro96}.
This is due, apart from the fundamental character of water, to the large 
variety of ice polymorphs which have been found under different conditions 
of pressure ($P$) and temperature ($T$).
The phase diagram of water is currently known up to $T \sim$ 1000 K and $P$
in the order of hundreds of GPa \cite{du10}.

Ice VII is generally accepted to be the stable phase of water at pressures 
from 2 to about 60 GPa at room temperature, although some studies predict
structural variations of ice VII in this pressure range \cite{su08,as10}.
Oxygen atoms in ice VII form a body-centered cubic (bcc) structure,
with H atoms lying between them, so that each O atom is surrounded by four
H atoms, two of which covalently bonded and two H-bonded to it,
according to the Bernal-Fowler rules \cite{be33}.
Apart from these constraints, the H distribution is disordered over
the available lattice sites \cite{ku84,sa11}.
The resulting H-bond network consists of two independent
interpenetrating subnetworks not hydrogen-bonded to each other, each
of them topologically equivalent to the H-bond network of cubic
ice Ic \cite{sa11,he13b,he14b}.

Ice VII is believed to be present in various natural environments, such as
the Earth mantle or icy planets. In this context,
the thermoelastic properties of ice VII and other high-pressure
polymorphs can have implications for the dynamics of cold slab subduction
in the Earth lower mantle, as well as for the evolution of icy planets
and satellites \cite{as10}.
This high-pressure water phase has been studied by a variety of techniques,
among which one finds X-ray \cite{he87,fe93,lo99,fr04,so08,su08,su10} and 
neutron \cite{ku84,ne98,kl99,kl09,fo12} diffraction, 
as well as vibrational spectroscopy \cite{wa82,kl84,so03,so03b},
Brillouin scattering \cite{as10},
electrical conductivity measurements \cite{li13}, and 
several kinds of calculations \cite{ho72,sl05,ve05b,su08,kn06,kn09,um10,fi13}.

At low temperatures, ice VII transforms into H-ordered ice 
VIII \cite{so03,so03b,du10}.
For $P <$ 10 GPa, the transition temperature $T_c$ is about 270 K,
and for $P >$ 10 GPa, $T_c$ decreases for increasing pressure \cite{so03b}.
At $P \sim$ 60 GPa, ice VII transforms into ice X, where H atoms lie
on the middle point between adjacent oxygen atoms, so that the molecular
nature of water is no longer preserved.
This phase transition from ice VII to ice X has been found to be caused
by quantum proton delocalization \cite{ka13,br14}.
At higher pressures, other ice phases have been encountered, and some
others are predicted to appear at pressures above 1 TPa \cite{he12}.

Computer simulation of condensed phases of water at an atomic level 
date back to the 1970s \cite{ba69,ra71}.
In the last few decades, much work has been devoted to the development 
of empirical potentials to describe both liquid and solid H$_2$O phases, 
so that nowadays a large variety of this kind of potentials can be found 
in the literature \cite{ma01,ko04,jo05,ab05,pa06,mc09}.
Many of these interatomic potentials assume a rigid geometry for the 
water molecule, and some others allow for molecular flexibility either 
with harmonic or anharmonic OH stretches. 
In recent years, simulations of water using \textit{ab initio} density 
functional theory (DFT) have been also carried out \cite{ch03,fe06,mo08}.
It has been observed, however, that a good description of H-bonds in 
condensed phases of water is a tough job with presently available energy
functionals, and some properties cannot be accurately 
reproduced by DFT calculations \cite{yo09}.
Some contributions to improve the description of van der Waals 
interactions in water within the DFT formalism 
have been developed in recent years \cite{le06,le07,wa11,ko11,ak11}.

A limitation of {\em ab-initio} electronic-structure calculations is that
they usually deal with atomic nuclei as classical particles, without
considering quantum aspects like zero-point motion.
Such quantum aspects can be taken into account by using harmonic
or quasiharmonic approximations for the nuclear motion, but the precision
of these methods is not readily assessed when large anharmonicities
are present, as may be the case for light atoms like hydrogen.
To consider the quantum character of atomic nuclei, the path-integral
molecular dynamics (or Monte Carlo) approach has turned out to be very
useful, because in this method the nuclear degrees of freedom can be
quantized in an efficient manner, including
both thermal and quantum fluctuations in many-body systems
at finite temperatures \cite{gi88,ce95}.
Then, a powerful method can consist in combining DFT to determine the
electronic structure and path integrals to describe the quantum motion of
atomic nuclei \cite{ma96b,tu96,ch03,mo08}. However, this procedure 
requires computer
resources that may enormously reduce the number of state points that
can be considered in actual simulations, as compared to the use of 
effective interatomic potentials.

Effective potential models for condensed phases of water usually treat
H$_2$O molecules as well-defined unbreakable entities, connected by
H-bonds. This prevents their use to study high-pressure phases such as
ice X, including symmetric O--H--O bonds.
This shortcoming does not affect {\em ab initio} potentials,
where in principle any bonding configuration or geometry can be
studied.  In this line, Benoit~{\em et al.} \cite{be98,be02}
investigated the influence of nuclear quantum motion
in proton ordering and H-bond symmetrization in high-pressure phases
of ice, by carrying out path-integral molecular dynamics (PIMD) 
simulations with an interatomic interaction given by DFT calculations.
More recently, the momentum distribution of protons in ice X was studied
by Morrone~{\em et al.} \cite{mo09} using also {\em ab initio} PIMD.

Nuclear quantum effects become more relevant for light atomic masses, 
and are expected to be especially important in the case of hydrogen.
Then, we consider the question of how quantum motion of the
lightest atom can influence the structural properties of a solid water
phase such as ice VII, and in particular if these effects are relevant or
detectable for the solid at different densities, i.e. under different
external pressures. This refers to the crystal volume and interatomic
distances, but also to the thermodynamic properties of ice. 
In this respect, it is interesting to estimate the effect of nuclear 
quantum motion on the stability range of the different ice phases.
Thus, it is known that quantum effects change the melting temperature 
$T_m$ of hexagonal ice Ih at ambient pressure by some degrees, as
manifested by the isotope effect on $T_m$ \cite{ra10}.

In this paper we study ice VII by PIMD simulations at several pressures 
and temperatures, in order to analyze its structural and thermodynamic 
properties. Interatomic interactions are described by the flexible 
q-TIP4P/F model, which was used earlier to carry out PIMD simulations of 
liquid water \cite{ha09}, ice Ih \cite{ra10,he11,he11b}, and 
high-density amorphous ice \cite{he12b}.
This potential model has been found to yield a reasonable description of 
the phase diagram of water up to pressures in the order of 
10~GPa \cite{ra13}.
The use of a rather accurate effective potential allows us to carry out
long PIMD simulation runs, and to obtain a good convergence of structural
and thermodynamic variables in the $P-T$ region under consideration.
The q-TIP4P/F potential model was found to yield reliable results for
low-pressure phases such as ice Ih \cite{ra11,he11,he11b}, and it is 
a goal of the present work to analyze its capability to describe 
high-pressure phases as ice VII.
A different goal consists in evaluating the reliability of applying a QHA 
to study these phases, by comparing its predictions with results of PIMD 
simulations. Such an approximation has been used earlier to study 
low-pressure water phases such as ices Ih, II, and III \cite{ra12}.

 The paper is organized as follows. In Sec.\,II, we describe the
computational methods and the model used in our calculations. 
Our results are presented in Sec.\,III, dealing with the pressure
and temperature dependence of molar volume, bulk modulus, O--H interatomic 
distance, kinetic energy, and atomic delocalization in ice VII.  
Sec.\,IV gives a summary of the main results.

\section{Computational Method}

\subsection{Path-integral molecular dynamics}

We use the PIMD method to obtain equilibrium properties of ice VII
at various temperatures and pressures.
This method is based on an isomorphism between the real quantum system
and a fictitious classical one, that appears after a discretization of 
the quantum density matrix along cyclic paths \cite{fe72,kl90}.
Such an isomorphism translates in practice into replacing each 
quantum particle by a ring polymer consisting of $L$ (Trotter number) 
classical particles, which are connected by harmonic springs with 
temperature- and mass-dependent force constant. 
This isomorphism is exact for $L \to \infty$, so that a numerical error 
is introduced by using a finite $L$, which may be corrected by adequate
extrapolation of the results to infinite $L$ (see below).
Details on this simulation technique can be found 
elsewhere \cite{gi88,ce95,he14}.
The dynamics considered in the PIMD method is artificial,
and does not correspond to the actual quantum dynamics of the real particles 
under consideration. It is, however, useful for effectively sampling the
many-body configuration space, giving precise results for the equilibrium
properties of the quantum system.
An alternative way to obtain equilibrium properties is 
Monte Carlo sampling, but this procedure requires for our 
present problem more computer resources (CPU time) than the PIMD method.
In particular, for the latter procedure the codes can be more readily 
parallelized, which is an important factor for efficient use of modern
computer architectures.

Ice VII has a body-centered cubic structure with space group $Pn3m$
and two water molecules per unit cell \cite{ku84}.
For the simulations we generated cubic supercells including
$N$ = 432 water molecules with periodic boundary conditions.
In these supercells, hydrogen-disordered ice structures were generated 
by a Monte Carlo procedure, in such a way that each oxygen atom had two 
covalently bonded and two H-bonded hydrogen atoms 
(Bernal-Fowler rules \cite{be33}).
It was also imposed that the electric dipole moment of the generated
supercells were close to zero \cite{bu98}.
To quantify the influence of hydrogen disorder on the results
presented below, we carried out PIMD simulations for six different
hydrogen arrangements. We found that the dispersion of the results found 
for the different H configurations was smaller than the statistical 
error bars obtained from simulations for a single H arrangement.  
This was taken as an indication that the supercell size considered
here was enough to adequately represent the proton disorder in ice
VII, in accord with earlier results for this ice phase \cite{ra13}.

Interatomic interactions have been modeled here by the point charge,
flexible q-TIP4P/F model, which was applied earlier to study liquid
water \cite{ha09,ra11}, ice \cite{ra10,he11,he11b}, and 
water clusters \cite{go10}.
Several empirical potentials previously used for 
simulations of condensed phases of water treat H$_2$O molecules as rigid 
bodies \cite{he05,mi05,he06b}. This may be convenient for computational
efficiency, giving reliable results for various properties of liquid water 
and ice, but they disregard the contribution of molecular
flexibility in the structure and dynamics of condensed phases of 
water \cite{ha09}, aspects which may be relevant for our present purposes.
In particular, a flexible water molecule allows us to study correlations 
between the intramolecular O--H distance and the geometry of intermolecular
H-bonds, known to appear in ice \cite{ho72,ny06}.
Also, the appreciable anharmonicity of the intramolecular O--H vibration 
is described by anharmonic stretches in the q-TIP4P/F potential, 
which allows to study changes in the intramolecular O--H distance as a 
function of temperature and pressure, as well as compare results for
classical and quantum simulations (see below). 

It has been reported as a limitation of the q-TIP4P/F effective potential
that it is unable to predict the stability region of the low-pressure
phase ice II \cite{ha11a}. This result was derived from free energy 
calculations based on classical simulations of ices Ih, II, and III, which
indicated that ice II was not the stable phase in any region of the
studied $P-T$ plane (down to $T$ = 200~K). It has been later reported that
ice II appears as the stable phase in a broad region of the phase diagram
when nuclear quantum effects are taken into account, basically due to
the lower zero-point energy of this water phase, as compared to ices Ih
and III \cite{ra12b}. This shows that path-integral simulations can help to
elucidate structural and thermodynamic aspects of different ice phases,
especially in cases where energy (or free energy) differences derived from
classical calculations turn out to be small.

Simulations of ice VII have been carried out here in the isothermal-isobaric
$NPT$ ensemble, which allowed us to find the equilibrium volume of
the solid at given pressure and temperature.
We used effective algorithms for performing PIMD simulations
in this statistical ensemble, as those described in the
literature \cite{ma96,tu98,ma99,tu02}.  In particular, 
we employed staging variables to define the bead coordinates, and
the constant-temperature ensemble was generated by coupling chains
of four Nos\'e-Hoover thermostats to each staging variable.
An additional chain of four barostats was coupled to the 
volume to yield the required constant pressure,\cite{tu98} which was 
computed along the simulations by using a virial expression adequate
for PIMD simulations (see Eq.~(50) in Ref.~\cite{he14}).

Sampling of the configuration space has been carried out at temperatures
between 50 K and 450 K, and pressures between 1 bar and 12 GPa.
In some particular cases, we performed simulations at $T$ = 25~K,
in order to check the low-temperature convergence of the data.
For comparison with results of PIMD simulations, some classical 
molecular dynamics (MD) simulations of ice VII have been also carried out.
This can be realized in our context by setting the Trotter number $L$ = 1.

Other technical details about the simulations presented here are the 
same as those employed and described earlier in \cite{he11,he11b}.
The Trotter number $L$ was taken proportional to the inverse temperature 
($L \propto 1/T$), so that $L \, T$ = 6000~K, which keeps roughly a constant 
precision in the PIMD results at different temperatures.  
The largest error associated to the finite Trotter number $L$ is caused 
by the vibrational modes with highest frequency, i.e., the intramolecular 
O--H stretching modes.
This has been corrected by carrying out some simulations with larger $L$,
and extrapolating results of the studied properties for $L \to \infty$
as explained elsewhere \cite{ra12}.
The results shown below are the extrapolated ones, after correcting the
error due to the finite $L$.
In the PIMD simulations,
the time step $\Delta t$ associated to the calculation of interatomic 
forces was taken in the range between 0.1 and 0.3 fs, which yielded
adequate convergence for the variables studied here.
For given temperature and pressure, a typical simulation run consisted of 
$10^5$ PIMD steps for system equilibration, followed by 
$6 \times 10^5$ steps for the calculation of ensemble average properties.

PIMD simulations can be used to obtain insight into the atomic delocalization
at finite temperatures. This includes a thermal (classical) delocalization,
as well as a delocalization associated to the quantum character of the
atomic nuclei, which is quantified by the spatial extension of the paths 
associated to a given atomic nucleus.
For each quantum path, one can define the center-of-gravity (centroid) as
\begin{equation}
   \overline{\bf r} = \frac{1}{L} \sum_{l=1}^L {\bf r}_l  \; ,
\label{centr}
\end{equation}
where ${\bf r}_l$ is the position of bead $i$ in the associated ring
polymer.
Then, the mean-square displacement $(\Delta r)^2$ of the atomic nuclei 
(H or O in our case) along a PIMD simulation run is defined as
\begin{equation}
  (\Delta r)^2 =  \frac{1}{L} \left< \sum_{l=1}^L 
           ({\bf r}_l - \left< \overline{\bf r} \right>)^2
           \right>    \, ,
\label{delta2}
\end{equation}
where $\langle ... \rangle$ indicates an ensemble average.

In connection with the quantum delocalization of a particle, 
a quantity related with its kinetic energy is the spread of the paths 
associated to it,
which can be measured by the mean-square ``radius-of-gyration''
$Q_r^2$ of the ring polymers:
\begin{equation}
  Q_r^2 = \frac{1}{L} \left< \sum_{l=1}^L
             ({\bf r}_l - \overline{\bf r})^2 \right>    \, .
\label{qr2}
\end{equation}
A smaller $Q_r^2$ (higher particle localization) is associated 
with a higher kinetic energy, in accordance with Heisenberg's 
uncertainty principle \cite{gi88,gi90}.

\subsection{Quasi-harmonic approximation}

We will compare below results of the PIMD simulations with those yielded
by a quasi-harmonic approximation (QHA), carried out by using the 
same q-TIP4P/F interatomic potential. 
This procedure has been introduced earlier in \cite{ra12} 
and \cite{ra12b}, and we will only present here a brief summary of it.

The Helmholtz free energy of a solid phase with $N$ water molecules in
a cell of volume $V$ and at temperature $T$ is given by
\begin{equation}
  F(V,T) = U_0(V) + F_{\rm v}(V,T) - T \, S_{\rm H}  \;,
\label{fvt}
\end{equation}
where $U_0(V)$ is the static zero-temperature classical energy,
i.e., the minimum of the potential energy for volume $V$. 
$F_{\rm v}(V,T)$ is the vibrational contribution to the free energy, 
which in a quantum approach is given by
\begin{equation}
  F_{\rm v}(V,T) = \sum_i \left( \frac12 \hbar \, \omega_i + 
         k_B T \ln \left[1 - \exp \left(- \frac{\hbar \, \omega_i}{k_B T}
        \right) \right] \right)  \;,
\label{fvv}
\end{equation}
where $\omega_i$ are the frequencies of 
the harmonic lattice vibrations for volume
$V$, with $i$ including the phonon branch index and the wave vector
within the Brillouin zone. The anharmonicity of the interatomic potential
enters in the QHA only through the volume dependence of $\omega_i$.
In the classical limit the vibrational contribution amounts to
\begin{equation}
  F_{\rm v,cla}(V,T) = \sum_i k_B T
        \ln \left( \frac{\hbar \, \omega_i}{k_B T} \right)  \;.
\label{fvc}
\end{equation}

$S_{\rm H}$ in Eq.~(\ref{fvt}) is the configurational entropy due to 
H disorder, and is relevant when comparing different ice phases, in 
particular H-ordered and H-disordered ones.
In our case, it can be taken as the Pauling entropy \citep{pa35} or that 
given by some more elaborate calculations \cite{he13,he14c},
but its actual value does not affect the results discussed here.

To obtain the Gibbs free energy, $G(T,P)$, one finds the volume 
$V_{\rm min}$ that minimizes the function $F(V,T) + P V$. Then, $G(T,P)$ 
is obtained as
\begin{equation}
  G(T,P) = F(V_{\rm min},T) + P V_{\rm min}  \;.
\label{gtp}
\end{equation}
More details on the implementation of this QHA to study structural and
thermodynamic properties of ice phases can be found 
elsewhere \cite{ra12,ra12b}.

\section{Results}

\subsection{Volume}

\begin{figure}
\vspace{-1.0cm}
\includegraphics[width= 8cm]{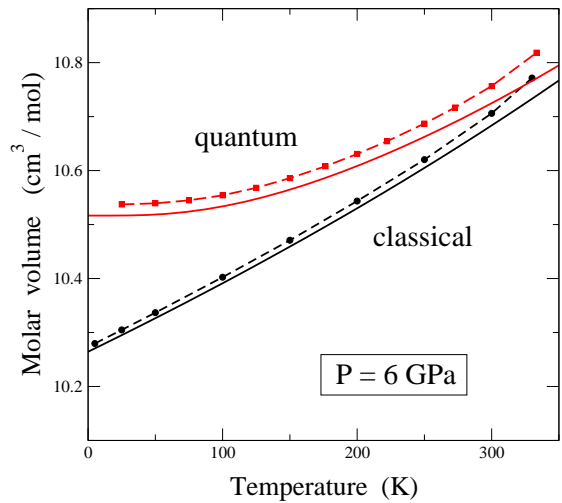}
\vspace{-0.3cm}
\caption{
Temperature dependence of the molar volume of ice VII for $P$ = 6 GPa.
Circles represent results of classical MD simulations, and squares are
data points derived from PIMD simulations.
Error bars are less than the symbol size.
Solid lines were obtained from classical and quantum QHAs.
}
\label{f1}
\end{figure}

We first present results for the equilibrium volume of ice VII, as derived 
from our PIMD simulations.
In Fig.~1 we show the temperature dependence of the molar volume at
a pressure $P$ = 6 GPa, in the temperature range where the solid was found
to be stable or metastable in the simulations (up to $T \sim$ 330 K).
Results of the quantum simulations are given as solid squares.
For comparison, we also present data points derived from classical MD
simulations (solid circles), as well as results of classical and quantum
QHAs (solid lines).
Looking at the results of the simulations, one observes the lattice
expansion due to quantum nuclear motion, which is most appreciable
at low temperatures.
At 25 K, we find $V$ = 10.305 cm$^3$/mol in the classical approach vs
10.537 cm$^3$/mol in the quantum simulations. This means a low-temperature
lattice expansion of 0.232 cm$^3$/mol, i.e., an increase of 2.3\% with
respect to the classical estimate.
This relative lattice expansion is somewhat smaller than that found
for ice Ih at low temperatures and ambient pressure ($\sim$ 4\%),
in agreement with the fact that it is expected to decrease as pressure
is raised \cite{he11,he11b}.

For ice VII at $P$ = 6 GPa, the quantum QHA gives values of the molar 
volume smaller than the PIMD simulations by about 0.02 cm$^3$/mol
(a difference of about 0.2\%),
and this difference increases slowly as the temperature is raised.
In the classical approach, both methods yield similar results at low 
temperatures, and the difference between them also increases for rising
$T$ as in the quantum case, since anharmonicity becomes more relevant.

\begin{figure}
\vspace{-1.0cm}
\includegraphics[width= 8cm]{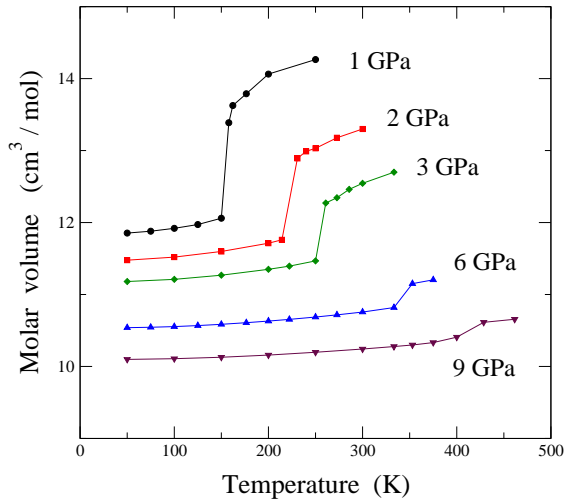}
\vspace{-0.3cm}
\caption{
Molar volume of ice VII as a function of temperature, as derived from
PIMD simulations at various pressures. From top to bottom: $P$ =
1, 2, 3, 6, and 9 GPa.  For pressures lower than 6 GPa, ice VII is
metastable in the temperature region shown here, up to the melting
of the crystalline solid displayed as a jump in the volume.
For $P$ = 6 and 9 GPa, ice VII is the stable phase
with the q-TIP4P/F potential up to $T \gtrsim$ 300 K \cite{ra13}.
Lines are guides to the eye.
}
\label{f2}
\end{figure}

An interesting aspect of Fig.~1 is the absence of a negative thermal 
expansion in the whole temperature range, contrary to what happens for 
low-pressure phases such as hexagonal ice Ih, and observed in PIMD 
simulations \cite{he11} as well as in diffraction experiments \cite{ro94}.
This negative expansion has been interpreted as being due to the
tetrahedral coordination of water molecules in ice Ih, along with the
large cavities (vacant space) present in its structure.
Although the H-bond network in ice VII has also a tetrahedral structure,
such cavities are not present in it because of the existence of two
subnetworks filling the empty space, with the associated increase in
density. 

In a recent study of the phase diagram of ice using the q-TIP4P/F potential,
it was found that ice VII is the stable phase only for pressures larger
than 6 GPa and $T \gtrsim$ 70 K \cite{ra13}. Thus, our results for $P <$ 6 GPa
correspond to metastable ice VII. In this pressure region, it is known 
that for a given value of $P$, ice VII becomes unstable at a
certain temperature that rises with $P$, and the crystalline solid
melts giving rise to an amorphous phase \cite{sl05,yo06}.
This behavior has been reproduced by our PIMD simulations. 
In Fig.~2 we present the molar volume of ice VII as a function of 
temperature at five different pressures. The three upper curves correspond 
to metastable ice VII, and their metastability 
limit is observed for each pressure as a jump in the volume. 
At 1 GPa, this jump occurs at about 150 K, and evolves to higher
temperatures as the pressure is raised. 
The two lower curves (for 6 and 9 GPa) correspond to the pressure region
where ice VII is the stable phase \cite{ra13}, and a jump in the molar 
volume is observed at $T \sim$ 350 and 400 K, respectively.

\begin{figure}
\vspace{-1.0cm}
\includegraphics[width= 8cm]{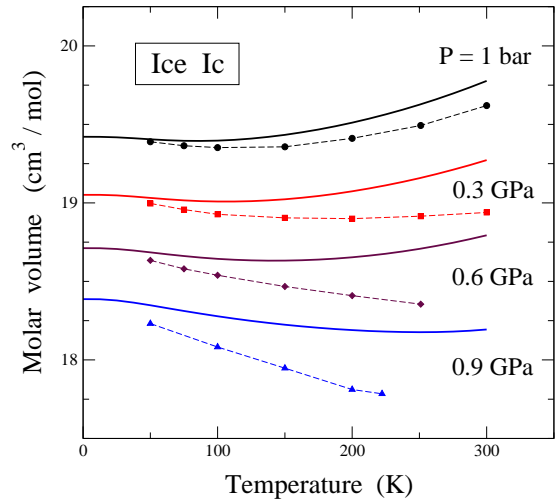}
\vspace{-0.3cm}
\caption{
Molar volume of ice Ic as a function of temperature at several pressures.
From top to bottom: $P$ = 1 bar, 0.3 GPa, 0.6 GPa, and 0.9 GPa.
Symbols represent results of PIMD simulations. The solid lines were obtained
from a quantum QHA.
}
\label{f3}
\end{figure}

Even though our main interest here is focused on ice VII, it is
worth commenting about the phases we observed in PIMD simulations
when ice VII became unstable by increasing the temperature.
According to the obtained radial distribution functions (RDFs),
for $P <$ 6 GPa we found an amorphous phase similar to those
described earlier by Slov\'ak and Tanaka \cite{sl05} from classical
molecular dynamics simulations using TIP4P and SPC/E potentials.
These authors explored the possibility of generating model structures 
of low-density amorphous ice by computationally heating ice VII.
For $P$ = 6 and 9 GPa, we found an ice structure, which according to 
the RDF is compatible with the plastic crystal phase detected earlier in 
various computational studies of ice in this pressure region, 
using different interatomic potentials \cite{ta08,hi11,ar09,ar09b}. 
It is remarkable that the RDF obtained
here from PIMD simulations using the q-TIP4P/F potential is very
similar to that reported earlier for the body-centered cubic plastic 
crystal \cite{ar09,ar09b}.

As mentioned in the Introduction, it is worthwhile comparing results of
PIMD simulations for a cubic high-pressure phase of water with those
derived for a cubic low-pressure phase such as ice Ic.
Thus, the temperature dependence of cubic ice Ic is shown in Fig.~3 for 
several pressures. Again, symbols represent results of PIMD simulations,
and solid lines were obtained in a quantum QHA.
At ambient pressure ($P$ = 1 bar), the molar volume displays a minimum at
about 120 K, similarly to that reported earlier for PIMD simulations
of hexagonal ice Ih at this pressure \cite{he11}.
In fact, the results for the molar volume of Ih and Ic ices derived from
the simulations coincide within error bars.
For increasing pressure, one observes that the minimum of the volume 
shifts towards higher temperatures, and eventually disappears when this
ice phase becomes unstable.
Results of the QHA present qualitatively the same trend for the molar 
volume of ice Ic as the PIMD data, but the former overestimates the negative 
thermal expansion.
This is a case where the PIMD and QHA results are appreciably different,
as a consequence of the instability appearing in ice Ic (as in ice Ih),
which causes the collapse of the crystal structure at a pressure of 
about 1 GPa \cite{mi84,he11b}.

\begin{figure}
\vspace{-1.0cm}
\includegraphics[width= 8cm]{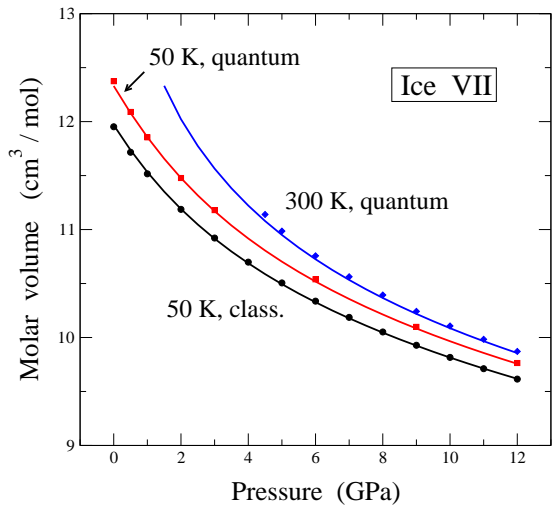}
\vspace{-0.3cm}
\caption{
Molar volume of ice VII as a function of pressure. Symbols indicate
simulation results: circles, classical MD at 50 K;
squares, PIMD at 50 K; diamond, PIMD at 300 K.
Error bars are less than the symbol size.
Solid lines represent results of QHAs for the same conditions
as the corresponding simulation data.
}
\label{f4}
\end{figure}

The negative thermal expansion of ice Ic shown in Fig.~3 is due to 
the negative Gr\"uneisen parameter $\gamma_i$ of low-energy transverse 
acoustic (TA) modes, related to librational motion of water molecules. 
For a vibrational mode with frequency $\omega_i$, the parameter
$\gamma_i$ is defined as the logarithmic derivative of $\gamma_i$ 
with respect to the crystal volume \cite{as76}:
\begin{equation}
  \gamma_i = - \frac {\partial \ln \omega_i} {\partial \ln V} \, .
\label{gamma}
\end{equation}
At relatively low temperature, low-energy modes are more populated than
modes with higher energy (with positive $\gamma_i$), and then
the overall contribution to the volume change with increasing
temperature will be negative \cite{he11}.
In this context, the larger negative thermal expansion of ice Ic derived
from the QHA, with respect to the results of PIMD, is due to an
overestimation of the negative value of $\gamma_i$ for TA modes in
the QHA. Such overestimation increases for rising pressure, as 
shown in Fig.~3.

Turning now to the high-pressure phase,
in Fig.~4 we show the pressure dependence of the molar volume of
ice VII, as obtained from simulations with the q-TIP4P/F potential.
Symbols indicate simulation results, whereas lines were derived from 
QHAs, both classical and quantum.
Two temperatures are shown: 50 K and 300 K.
At 50 K, both classical and quantum data are given, and at 300 K only
the quantum results are shown (the classical ones lie slightly below
the quantum data, and are not displayed to avoid overcrowding  
of the figure).

The quasi-harmonic approximation gives a good description of the equation 
of state $P-V$ of ice VII in both cases, quantum and classical.
At 300 K, the PIMD simulations yield volumes slightly higher than 
the QHA, but the difference between both methods decreases as the pressure is
raised. This is in line with earlier observations that the QHA
becomes more accurate to describe structural properties of solids 
under high pressures \cite{po72}.
This is related to the pressure (or volume) dependence of the vibrational 
$F_{\rm v}$ and elastic $U_0$ contributions to the free energy in 
Eq.~(\ref{fvt}).
For high pressures (small volumes), the internal energy is dominated
by the elastic contribution $U_0$, which becomes much larger than 
the vibrational energy, and hence the influence of lattice vibrations
(and of their anharmonicity) on thermodynamic properties comparatively
decreases \cite{he05d}.
This means that the anharmonicity of the vibrational modes 
not captured by the QHA becomes less relevant for high pressures.
On the contrary, we emphasize that the relative error of the QHA 
(compared with PIMD simulations) increases as temperature is raised, 
as expected for an increased 
anharmonicity due to thermal effects and shown in Fig.~1.
This difference between QHA and PIMD usually increases slowly for 
rising $T$, unless some particular anomaly appears for the vibrational
frequencies, as happens close to the stability limit of a solid phase
(see Fig.~3 for ice Ic, which becomes unstable for $P \sim$ 1~GPa).
In such a case, the QHA also becomes less accurate for increasing $P$,
contrary to the general trend discussed above.

One also observes in Fig.~4 that the difference between quantum and 
classical results at 50 K decreases as $P$ rises.
This difference amounts to 0.42 cm$^3$/mol for $P$ = 1 bar vs
0.15 cm$^3$/mol for $P$ = 12 GPa, i.e., the quantum-induced volume
expansion decreases from a 3.5\% to a 1.6\% in this pressure range. 
This is in line with the fact that the classical model become more and
more accurate to describe structural and thermodynamic properties
of solids at high pressures. 
This could seem at first sight contradictory with the fact that pressure 
induces a larger zero-point vibrational energy of the solid, of purely
quantum nature, because of the overall increase of vibrational frequencies
with rising pressure. However, the relevant factor here is again 
(as for the QHA) the ratio of the vibrational energy to the whole internal 
energy of the solid.
Since in this respect the lattice vibrations become less relevant as 
pressure rises, and eventually give a relatively small contribution
to the free energy of the solid, their actual description by a classical
or a quantum model becomes less important for solids under large 
pressures \cite{et74,he05d}.
This is not necessarily true for spectroscopic properties of solids,
since vibrational frequencies predicted by a QHA or by
a classical model are not guaranteed to describe correctly the actual ones
for high pressure.
Moreover, the above considerations on structural and thermodynamic 
properties apply to isolated phases such as ice VII here, but dynamical 
quantum effects can be crucial to describe phase transitions between 
different ice phases, as happens for H-bond symmetrization in the 
transition from ice VII to ice X \cite{be98,ka13,br14}.

\begin{figure}
\vspace{-1.0cm}
\includegraphics[width= 8cm]{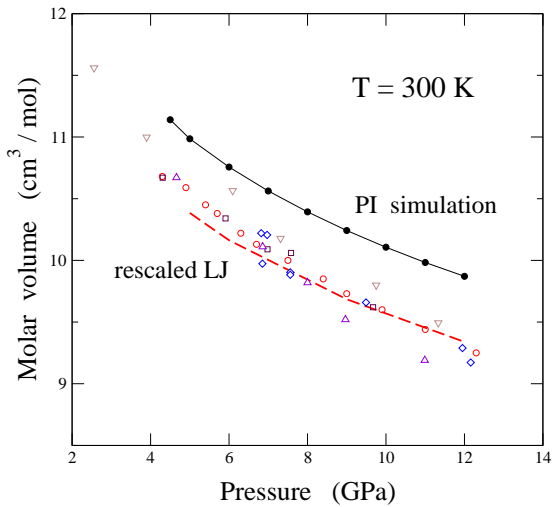}
\vspace{-0.3cm}
\caption{Molar volume of ice VII vs pressure at 300 K. Solid circles
represent results of PIMD simulations with the q-TIP4P/F potential.
Open symbols indicate data derived from diffraction experiments by
different authors:
circles \cite{he87}, squares \cite{fe93}, diamonds \cite{fr04},
triangles up \cite{wo97}, and triangles down \cite{lo99}.
The dashed line corresponds to results of PIMD simulations of ice VII with
the Lennard-Jones (LJ) interactions between oxygen atoms in different
subnetworks rescaled by a factor of 0.6.
}
\label{f5}
\end{figure}

An interesting question related to our simulations of H-disordered
ice VII refers to the dispersion in calculated volumes for different
realizations of H disorder on the simulation supercell.
As indicated in Sec.~II.A, we have considered six realizations for
the H distribution according to the ice rules, and differences between
the resulting volumes were less than the symbol sizes in Figs.~2 and 4. 
Thus, at 50 K our PIMD simulations yielded volume differences in the
order of $10^{-3}$ cm$^3$/mol. We note that this value is much smaller 
than the volume difference between ice VII and its H-ordered
counterpart ice VIII, for which we found from PIMD simulations 
a molar volume of 12.333 cm$^3$/mol at 1 bar and 50 K, i.e., 
0.041 cm$^3$/mol less than that found for ice VII a the same conditions.

Most measurements of the $P-V$ curve for ice VII have been 
performed at or close to room temperature.
In Fig.~5 we display the results of our PIMD simulations for
the molar volume at 300 K, along with data derived by several authors 
from X-ray diffraction experiments. At this temperature,
ice VII turns out to be unstable in PIMD simulations with the q-TIP4P/F
potential at pressures $P \lesssim$ 4~GPa, so that 4.5 GPa is the lowest 
pressure plotted in this figure for the simulation data.
Although there is some dispersion in the diffraction data, the molar 
volumes obtained from experiment are smaller than those predicted
by the simulations. The difference between calculated and actual
volumes decreases as pressure is reduced, and both seem to converge
at low pressures. This could be expected from the good agreement 
obtained for low-pressure phases (ices Ih, II, and III), between
measured molar volumes and those derived from PIMD simulations
with the q-TIP4P/F potential model \cite{ra12}.

At 4.5 GPa the simulations overestimate the molar volume of ice VII
by about a 4\%, and this difference increases as pressure grows,
reaching a 6\% at 12 GPa.
We have investigated the origin of this drawback of the q-TIP4P/F
potential model at high pressures, and found that the solid
becomes ``stiff'' mainly due to the Lennard-Jones-type contribution to 
the interaction between adjacent molecules in different subnetworks. 
To check this,
we have carried out some PIMD simulations with a modified q-TIP4P/F
potential where this contribution was rescaled by a factor less
than one, so that the repulsion between water molecules in both
subnetworks was reduced. 
The dashed line in Fig.~5 corresponds to a simulation with such a 
factor taken to be 0.6, which means that the Lennard-Jones interaction 
between different H-bond subnetworks was reduced by a 40\%. 
With such a correction in the interatomic potential, the simulation 
results are much closer to those derived from diffraction data, 
especially at pressures higher than 6 GPa. 
This shows a limitation of this kind of interatomic potentials for
studying water phases at very high pressures. An {\em ad hoc} modification
of the potential may improve the description of some features
of these phases, but it is not clear that it can give a good overall 
description of their structural and thermodynamic properties. 
A similar conclusion was drawn by Aragones {\em et al.} \cite{ar09}
from their classical Monte Carlo simulations with a
potential model of rigid water. 

We finally note in this Section, that the results found for the molar volume
of ice VII using classical molecular dynamics simulations with the
q-TIP4P/F potential are close to those obtained earlier employing the
TIP4P/2005 model.
Thus, we have found at 300 K and pressures of 4.5 and 7 GPa, molar volumes
$v$ = 11.060 and 10.514 cm$^3$/mol, respectively. These values are
near those reported in Ref.~\cite{ar09} for the TIP4P/2005 potential
($v$ = 11.077 and 10.545 cm$^3$/mol), and higher than those found for 
the TIP5P potential model (10.418 and 9.923 cm$^3$/mol).
A similar conclusion is drawn from a comparison of zero-pressure results
at 70 K with those given in Ref.~\cite{sl05}. For these conditions, we
find a classical result for the molar volume of 12.036 cm$^3$/mol, clearly
higher than the value found for the TIP5P potential (11.288 cm$^3$/mol),
and closer to the SPC/E (11.931 cm$^3$/mol) and TIP4P results 
(11.876 cm$^3$/mol).

\subsection{Bulk modulus}

The compressibility $\kappa$ of ice phases displays peculiar properties 
related to their H-bond networks. 
For low-pressure phases, $\kappa$ is smaller than what one 
could at first sight expect from the large cavities present in their
structure, which could be supposed to collapse under pressure without 
water molecules approaching close enough to repel each other.
This is however not the case for low pressure, and for ice Ih the hydrogen 
bonds holding the crystal structure are stable up to relatively high 
pressures in the order of 1 GPa, when the H-bond network breaks down
giving rise to high-density amorphous ice \cite{mi84}.
For high-pressure ice phases such as ice VII, the compressibility
has been studied by several authors in its stability region.
In this case, the presence of two H-bond networks stabilizes the structure
up to pressures of about 60 GPa, where the H-bonds become symmetric 
and the transition to ice X occurs.

\begin{figure}
\vspace{-1.0cm}
\includegraphics[width= 8cm]{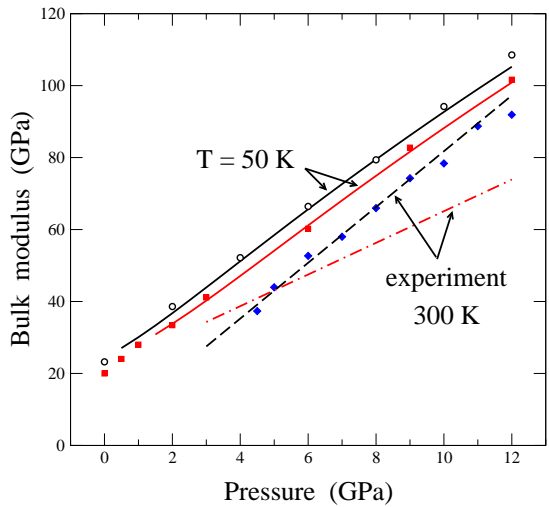}
\vspace{-0.3cm}
\caption{Bulk modulus of ice VII as a function of pressure.
Solid symbols represent results of PIMD simulations: squares, 50 K;
diamonds, 300 K. Open circles indicate results of classical MD simulations
at 50 K. Error bars are in the order of the symbol size.
Continuous lines were obtained from classical and quantum QHAs at 50 K.
Dashed and dashed-dotted lines represent data derived from X-ray diffraction
measurements by Loubeyre {\em et al.} \cite{lo99} and
Frank {\em et al.} \cite{fr04}, respectively.
}
\label{f6}
\end{figure}

The isothermal compressibility $\kappa$ of ice, or its inverse the bulk
modulus [$B = 1/\kappa = - V ( {\partial P} / {\partial V} )_T$] can be
straightforwardly obtained from our PIMD simulations in the 
isothermal-isobaric ensemble. 
Calling ${\cal V}$ the volume of the ice supercell along a simulation
run, the isothermal bulk modulus can be calculated from the mean-square 
fluctuations of ${\cal V}$,
$\sigma_V^2 = \langle {\cal V}^2 \rangle - \langle {\cal V} \rangle^2$,
by employing the expression \cite{la80,he08}
\begin{equation}
    B = \frac{k_B T \langle {\cal V} \rangle}{\sigma_V^2}   \; .
\label{bulkm}
\end{equation}
Note that in our notation one has $V = \langle {\cal V} \rangle$.
Eq.~(\ref{bulkm}) has been used earlier to obtain the bulk modulus
of different types of solids from path-integral
simulations \cite{he00c,he08,he11b}.

In Fig.~6 we present data of the isothermal bulk modulus $B$ of ice VII
as a function of pressure. Symbols indicate results of simulations
with the q-TIP4P/F potential. As in Fig.~4 data are shown for PIMD
simulations at 50 K (solid squares) and 300 K (solid diamonds), 
as well as for classical MD simulations at 50 K (open circles). 
Solid lines represent results of classical and quantum QHAs at 50 K. 
These lines follow closely the results of the corresponding simulations, 
classical and quantum, in the pressure region considered here.  
From a linear fit of the PIMD results, we find a pressure derivative
of the bulk modulus: $B' = \partial B / \partial P$ = 6.8(1) and 7.2(2),
at 50 and 300 K, respectively. 
The dashed and dashed-dotted lines indicate the pressure dependence of 
the bulk modulus at 300 K, as derived from the equation of state 
$P-V$ obtained in \cite{lo99} and \cite{fr04} from 
X-ray diffraction measurements.
The results by Loubeyre {\em et al.} \cite{lo99} 
(zero-pressure bulk modulus: $B_0$ = 4.26 GPa; 
pressure derivative: $B_0'$ = 7.75) lie close to those obtained
from the quantum simulations, but other data such as those by 
Frank {\em et al.} \cite{fr04} ($B_0$ = 21.1 GPa, $B_0'$ = 4.4; 
dashed-dotted line) are smaller at pressures higher than 6 GPa.

\begin{figure}
\vspace{-1.0cm}
\includegraphics[width= 8cm]{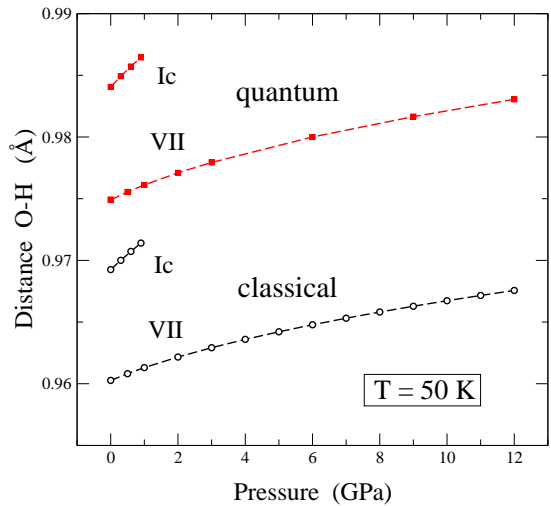}
\vspace{-0.3cm}
\caption{
Intramolecular O--H distance vs pressure at 50 K, as derived from
simulations with the q-TIP4P/F potential. Open and solid symbols represent
results of classical and PIMD simulations, respectively. Shown are data
for ices Ic and VII, as indicated by the labels.
Lines are guides to the eye.
}
\label{f7}
\end{figure}

We note that values of $B$ calculated from PIMD simulations in the 
isothermal-isobaric ensemble display relative error bars larger than 
those obtained for other variables (e.g., molar volume, kinetic energy, 
or interatomic distances), due to the statistical uncertainty 
in the volume fluctuations $\sigma_V$, used to calculate the bulk modulus.
An alternative way to obtain $B$ can consist in calculating numerically 
the derivative $\partial V / \partial P$ from the $P-V$ curve at a
temperature $T$. This procedure yields results that coincide within error 
bars with those found from the volume fluctuations $\sigma_V^2$ 
using Eq.~(\ref{bulkm}).

\subsection{Interatomic distances}

In this section we present results for interatomic distances in ice VII.
This can shed light on the structural changes suffered by the 
crystal when temperature and/or pressure are modified.
In Fig.~7 we show the mean O--H distance in ice VII as a function of 
the applied pressure at $T$ = 50 K.
Solid symbols represent results of PIMD simulations, whereas open circles
indicate data obtained from classical MD simulations. 
For comparison we also display results for ice Ic at the same temperature,
in the pressure region where this phase is stable (1 bar $\leq P \leq$ 1 GPa).
One observes first that for a given pressure, the distance O--H derived 
from the quantum simulations is larger than that found in the classical ones.
For ice VII this difference is roughly constant in the pressure region under
consideration, and amounts to 0.015 \AA\ (1.5\% of the bond length).
This bond expansion due to quantum motion is similar to that found for
ice Ih \cite{he11b}.

Second, one finds that the O--H distance increases as pressure is
raised. This seems in principle to be opposite to the usual contraction
of atomic bonds under hydrostatic pressure, and could also seem 
surprising given the pressure-induced volume contraction.
This is, however, characteristic of ice structures with hydrogen bonds
connecting water molecules, and can be explained as follows.
Increasing the pressure reduces the crystal volume, i.e. the interatomic
distance between adjacent oxygen atoms decreases in the bcc lattice of
ice VII. This causes a strengthening of intermolecular H-bonds, 
which is associated to a weakening of the intramolecular bonds, and
with an increase in interatomic O--H distance in water molecules.
An additional argument in favor of this reasoning is the softening of
O--H stretching modes as pressure increases, observed in infrared
absorption experiments on ice VII \cite{kl84,so03}.

\begin{figure}
\vspace{-1.0cm}
\includegraphics[width= 8cm]{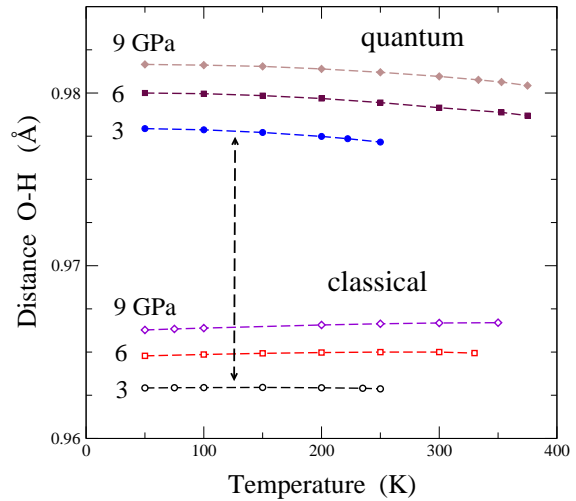}
\vspace{-0.3cm}
\caption{
Intramolecular O--H distance in ice VII vs temperature at three pressures:
$P$ = 3, 6, and 9 GPa. Open and solid symbols indicate results of
classical and PIMD simulations, respectively.
A vertical dashed line indicates the bond expansion due to nuclear quantum
motion for $P$ = 3 GPa.  Other lines are guides to the eye.
Error bars are smaller than the symbol size.
}
\label{f8}
\end{figure}

Comparison of results for ices VII and Ic shows that the O--H bond
expansion due to quantum nuclear motion is similar in both cases.
The intramolecular O--H bond distance is larger for ice Ic than for ice VII,
which is associated with a weaker covalent bond in ice Ic.
The origin of this is the following.
The molar volume of ice Ic obtained from PIMD simulations at 50 K and
1 bar amounts to 19.39 cm$^3$/mol, much larger than the result
for ice VII for the same pressure and temperature, $v$ = 12.09 cm$^3$/mol
(as expected for this high-pressure phase).
In spite of this, the intermolecular O--O distance is larger in ice VII
than in ice Ic, because
the diamond-like structure of ice Ic contains large void regions, which
are occupied by water molecules in ice VII.
We found in the PIMD simulations with the above conditions: 
$d$(O--O) = 2.99 \AA\ for ice VII vs 2.76 \AA\ for ice Ic.
Thus, larger O--O distances in ice VII imply weaker H-bonds
between water molecules, stronger intramolecular O--H bonds, and
therefore smaller O--H distances, as observed in Fig.~7.
This is in line with the local intra-intermolecular geometric correlation 
found for liquid and solid water, relating the intramolecular O--H bond 
length to the corresponding H-bond geometry \cite{ho72,ny06}.

One also observes in Fig.~7 that the change of O--H distance with pressure
is clearly larger for ice Ic, in accordance with a larger compressibility
of this low-pressure phase. In fact, the bulk modulus $B$ of Ic is found to 
be 14.2 GPa vs 20.0 GPa for ice VII (at 1 bar and 50 K).
In the range from 1 bar to 12 GPa we find for ice VII a mean rise in 
O--H bond distance of $7 \times 10^{-4}$ \AA/GPa. 
However, for pressures smaller than 2 GPa, a larger increase rate of
about $1.2 \times 10^{-3}$ \AA/GPa is obtained.
These values are smaller than those estimated from changes of stretching 
frequencies in \cite{kl84} ($1.8 \times 10^{-3}$ \AA/GPa).

We now turn to the temperature dependence of the O--H distance in ice VII.
In Fig.~8 we show the mean O--H distance for three pressures 
($P$ = 3, 6, and 9 GPa), as derived from classical (open symbols) and 
quantum (solid symbols) simulations.
For the classical results at $P$ = 3 GPa, we find that $d$(O--H) is rather 
constant up to 250~K, the highest temperature at which we found this ice 
phase to be metastable. For $P$ = 6 and 9 GPa, $d$(O--H) is found to
increase slightly with rising temperature. 
For the PIMD results, however, we find that the O--H distance decreases
for increasing temperature in the three cases shown in Fig.~8, so that
the difference between classical and quantum results is slowly reduced
as temperature rises.

The quantum results display clearly the anomalous trend observed for 
various phases of ice, for which the covalent O--H bond in water 
molecules contracts for increasing temperature.
The argument is similar to that given above for the pressure dependence
of the intramolecular O--H distance.
Given a pressure $P$, for rising $T$ the molar volume and the mean O--O 
distance increase, so that H-bonds between contiguous molecules become
weaker. The covalent O--H bonds are then stronger and the intramolecular 
O--H distance is smaller.  
This behavior is not captured by the classical MD simulations of ice VII. 
In fact, there is a competition between the usual trend of bond distances 
to increase for rising $T$ (in both quantum and classical models), and 
the trend to increase because of the weakening of the O--H bond.
In the classical simulations, the former effect dominates due to the
smaller atomic delocalization, whereas in the quantum model the latter
effect is more relevant.

There have appeared in the literature some data for the covalent
O--D bond length in high-pressure phases of D$_2$O, derived from
neutron diffraction measurements.
For ice VIII, this length is similar to that yielded by our simulations,
i.e., $d$(O-D) between 0.97 and 0.98 \AA\ for $P$ 
from 2 to 10 GPa \cite{ne93,be94}.   For D$_2$O ice VII, this distance 
was found to be smaller by Klotz~{\em et al.} \cite{kl99}   
Thus, at 98 K these authors found $d$(O-D) in the range from 0.92 to 
0.93 \AA\ for pressures between 2 and 6 GPa.  In ice VII, however, 
the presence of H-disorder (and the associated oxygen displacements) 
causes some uncertainties in the refinement of the diffraction data, 
which can be analyzed in different ways \cite{ne98}.

\begin{figure}
\vspace{-1.0cm}
\includegraphics[width= 8cm]{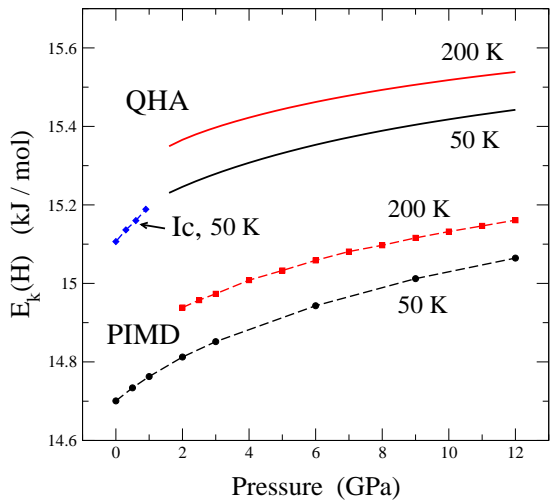}
\vspace{-0.3cm}
\caption{Kinetic energy of hydrogen in ice VII as a function of pressure
at 50 K and 200 K.
Symbols represent results of PIMD simulations, and the continuous lines
were obtained from a QHA approximation.
Diamonds represent the kinetic energy $E_k(H)$ for ice Ic, as derived from
PIMD simulations at 50 K and pressures between 1 bar and 1 GPa.
Error bars are less than the symbol size.
}
\label{f9}
\end{figure}

\subsection{Kinetic energy}

In this section we discuss the kinetic energy of hydrogen in ice VII.
First, we note that the kinetic energy of atomic nuclei in molecules 
or solids depends on the mass and spatial delocalization of the considered 
nucleus.  In a classical model, each degree of freedom 
contributes to the kinetic energy by an amount proportional to the 
temperature, $k_B T / 2$, as granted by the equipartition principle. 
In a quantum approach the kinetic energy, $E_k$, provides us with 
information on the environment and interatomic interactions seen by the 
particle under consideration. 
A common quantum effect associated to atomic motion in solids is
that the kinetic energy at low temperature converges to a finite value
characteristic of zero-point motion, in contrast to the classical result 
that $E_k$ vanishes for $T \to 0$ K.
Path integral simulations allow one to calculate the kinetic energy
of quantum particles, which is related to the spread of the quantum paths. 
In fact, for a particle with a certain mass at a given temperature, 
the larger the mean-square radius-of-gyration, $Q_r^2$ [see Eq.~(\ref{qr2}], 
the smaller the kinetic energy, since larger quantum delocalization 
yields a reduction in $E_k$ \cite{gi88,he11}.
Here we have calculated $E_k$(H) in ice VII by using the so-called virial 
estimator, which is known to have a statistical uncertainty appreciably 
smaller than the potential energy of the system \cite{he82,tu98}.

In Fig.~9 we present $E_k$(H) as a function of pressure at $T$ = 50 
and 200 K. 
Symbols (circles and squares) correspond to results of PIMD simulations, 
and solid lines represent the kinetic energy derived from the QHA.
In both cases, the results at 200 K are higher than those found at
50 K, as expected from the usual thermal activation of vibrational modes.
For a given temperature, $E_k$(H) increases as pressure is raised,
in agreement with an overall increase of vibrational frequencies
for increasing pressure.  

We also show in Fig.~9 the kinetic energy of hydrogen in ice Ic, as obtained
from PIMD simulations at 50 K. It is higher than that corresponding to
ice VII at the same temperature in the range from 1 bar to 1 GPa.
This is mainly due to the weakening of H-bonds in ice VII, as compared
with ice Ic, since the O--O distance is larger in ice VII (see above).
The contribution of O--H stretching modes to $E_k$(H) is larger in
ice VII (stronger covalent bonds, with larger zero-point energy), 
but it is largely compensated for by the contribution of librational 
modes (weaker H-bonds in ice VII).
We note that the QHA gives for ice Ic at $T$ = 50 K and $P$ = 1 bar
a kinetic energy $E_k$(H) = 15.47 kJ/mol (not shown in the figure),
i.e., a value somewhat larger than
that predicted by this approximation for ice VII at low pressure, in
line with the results of PIMD simulations for both ice phases.

We note that the actual momentum distribution of quantum particles can
be derived from atomistic simulations using open path integrals, 
in contrast with the closed paths employed here to specify the
statistical partition function \cite{fe72}. In this context, an
estimator for the end-to-end distribution of the Feynman paths was
introduced by Lin {\em et al.} \cite{li10}, giving the Fourier
transform of the momentum distribution.

\begin{figure}
\vspace{-1.0cm}
\includegraphics[width= 8cm]{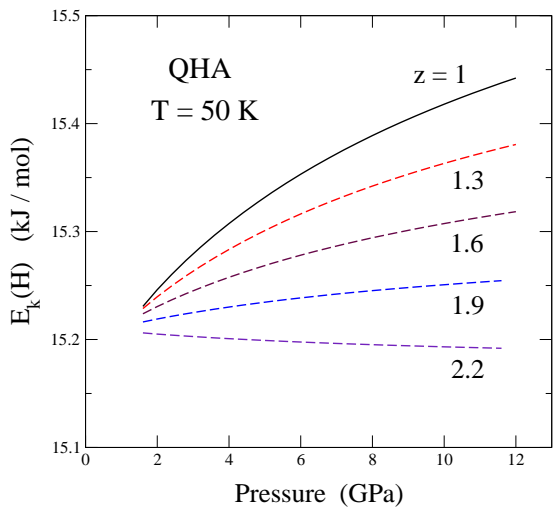}
\vspace{-0.3cm}
\caption{Kinetic energy of hydrogen in ice VII as a function of pressure
at 50 K, as derived from the QHA.
The solid line corresponds to the q-TIP4P/F potential, as in Fig.~9.
The dashed lines were obtained by rescaling the Gr\"uneisen parameters
$\gamma_i$ of the stretching O--H vibrations by different factors,
as indicated by the labels ($z$ = 1.3, 1.6, 1.9, and 2.2).
}
\label{f10}
\end{figure}

In connection with the pressure dependence of the kinetic energy of
hydrogen in ice VII, Finkelstein and Moreh \cite{fi13} have recently
presented a calculation based on a harmonic approximation with vibrational
frequencies taken from experimental data at $T$ = 300 K and various 
pressures.  In these calculations, $E_k$(H) was found to decrease 
in a large range of pressures, in particular in the region between
5 and 12 GPa, which overlaps with our calculations.
The main reason for this decrease is the softening of the O--H stretching
modes for rising pressure.
These calculations involved five representative frequencies for the
various vibrational modes present in ice VII (translation, libration,
bending, symmetric and asymmetric stretching), and not the whole 
vibrational density of states of the solid, but one expects
that the main features of the vibrational energy should be 
captured by such a model.

This behavior of $E_k$(H) is associated to the decrease of stretching
frequencies as pressure is raised, discussed above in connection with
the pressure-dependence of the molar volume.
Such frequency changes as a function of pressure (or volume) are usually 
quantified by means of the Gr\"uneisen parameters $\gamma_i$, defined 
above in Eq.~(\ref{gamma}).
In most solids, these parameters are usually positive, but in several solid 
phases of water $\gamma_i$ for TA vibrational modes, related to 
librational motion, and O--H stretching modes are known to be negative.
As indicated above, the negative values of $\gamma_i$ for TA modes 
in ice Ic, are related to the negative thermal expansion shown in Fig.~3.
Moreover,
a negative value of $\gamma_i$ for the stretching modes causes a decrease 
in $E_k$(H) for ice VII, as that calculated in \cite{fi13}. 
The q-TIP4P/F potential gives a negative Gr\"uneisen parameter for the 
stretching vibrations \cite{pa12} ($\gamma_i \sim -0.2$), but it is not 
negative enough to reproduce the inverse isotope effect in ice Ih \cite{he11},
which has been found from diffraction experiments \cite{ro94}, i.e.,
the molar volume of D$_2$O ice is larger than that of H$_2$O ice.

To understand the effect of a negative value of $\gamma_i$ for
stretching modes on the kinetic energy of hydrogen, we have calculated
$E_k$(H) in the QHA using the vibrational density of states derived from
the q-TIP4P/F potential, but with the stretching frequencies rescaled by
different factors.
From the definition of the Gr\"uneisen parameter $\gamma_i$ for a
mode $i$ [see Eq.~(\ref{gamma})], one has for small volume changes 
($\Delta V / V_0 \ll 1$):
\begin{equation}
 \frac{\Delta \omega_i}{\omega_{i0}} = - \gamma_i \, \frac{\Delta V}{V_0}
   \; ,
\label{deltaom}
\end{equation}
where $\omega_{i0}$ is the vibrational frequency for the reference 
volume $V_0$.
Thus, we have recalculated the frequency $\omega_i$ of the stretching
modes by using Eq.~(\ref{deltaom}) with a renormalized 
$\gamma_i' = z \, \gamma_i$, for several values of the parameter $z$
between 1 and 2.2. The other modes are left unaltered, with the
frequencies obtained from the q-TIP4P/F potential. 

In Fig.~10 we display the dependence of $E_k$(H) on pressure, as given
by the QHA at $T$ = 50 K. The solid line with a label ``$z = 1$'' 
coincides with the curve shown in Fig.~9 for this temperature.
The four other curves have been obtained by introducing in the 
QHA the stretching frequencies $\omega_i$ obtained from Eq.~(\ref{deltaom})
by scaling $\gamma_i$ with the factors indicated by the labels
($z$ = 1.3, 1.6, 1.9, ad 2.2). 
Since $\gamma_i$ are negative for the stretching modes,
this correction makes them more negative, thus favoring a faster 
decrease of the corresponding frequencies as the volume is reduced 
(or pressure is raised). 
For $z \gtrsim 2$, $E_k$(H) is found to decrease for rising pressure,
similarly to the results presented in \cite{fi13}.
This indicates the crucial role of stretching modes for the
pressure dependence of the vibrational energy. A large contribution of
these vibrational modes to the kinetic energy could be expected from 
their high frequency, as compared to the other modes in this solid.
The pressure-dependence of $E_k$(H) is however largely controlled by
the parameters $\gamma_i$, which turn out to be negative for the stretching
modes in condensed phases of water. Such a dependence is a critical check
for interatomic potentials in this kind of molecular solids.

\begin{figure}
\vspace{-1.0cm}
\includegraphics[width= 8cm]{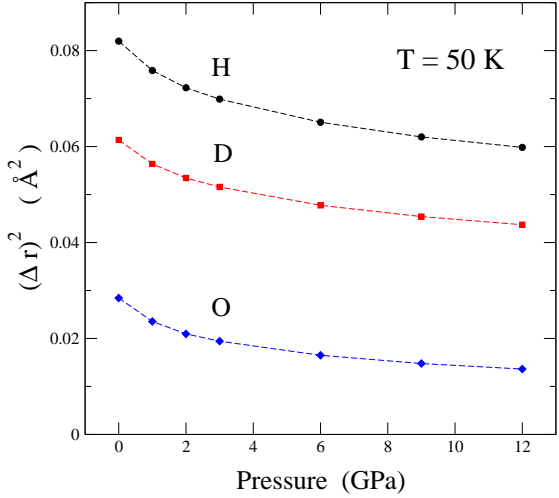}
\vspace{-0.3cm}
\caption{Mean-square displacement $(\Delta r)^2$ of H, D, and O atoms
in ice VII vs pressure at 50 K.
}
\label{f11}
\end{figure}

\subsection{Atomic delocalization}

Here we present results for the atomic mean-square displacements 
$(\Delta r)^2$ defined in Eq.~(\ref{delta2}).
The spatial delocalization of atomic nuclei in ice VII is expected 
to be larger for hydrogen than for oxygen, because of the smaller mass 
of the former atom.
In Fig.~11 we show values of $(\Delta r)^2$ for hydrogen and oxygen 
at 50 K as a function of pressure. 
At a given pressure, $(\Delta r)^2$ decreases as the pressure is raised,
i.e., both H and O atoms become more spatially localized.
For H, $(\Delta r)^2$ decreases from $8.2 \times 10^{-2}$ \AA$^2$ for
$P$ = 1 bar to $6.0 \times 10^{-2}$ \AA$^2$ for 12 GPa, which means a 
reduction of a 37\% in this pressure region.
For oxygen, $(\Delta r)^2$ decreases from $2.8 \times 10^{-2}$ \AA$^2$ 
to $1.4 \times 10^{-2}$ \AA$^2$, i.e., by a factor of 2. 

Results of $(\Delta r)^2$ of H and O in ice VII at ambient pressure are 
similar to those found earlier for hexagonal ice Ih at the same conditions.
Thus, for ice Ih at 50 K and 1 bar, it was found 
$(\Delta r)^2 = 7.9 \times 10^{-2}$ \AA$^2$ for hydrogen and
$4.1 \times 10^{-2}$ \AA$^2$ for oxygen \cite{he11b}.
For this low-pressure ice phase, however, it was found that rising
the pressure causes an increase in $(\Delta r)^2$ for both H and O, 
contrary to the decrease
found here for ice VII at the same temperature. This behavior for ice Ih
is indeed associated to the destabilization of the crystal structure,
which breaks down at a pressure $P \sim$ 1 GPa, with the consequent 
amorphization of the solid. For cubic ice Ic we have found values of
$(\Delta r)^2$ coinciding within error bars with those corresponding to 
ice Ih, and also the same behavior has been found in both ice phases 
for increasing pressure. 
No such anomalies are found for the atomic delocalization in
ice VII in the whole pressure region studied here.

In general, given a pressure, $(\Delta r)^2$ is expected to increase as 
temperature is raised. Such an increase has been obtained in the whole 
region where ice VII was found to be stable or metastable 
in our PIMD simulations.
Thus, at $P$ = 6 GPa, $(\Delta r)^2$ for H increases from 
$6.5 \times 10^{-2}$ \AA$^2$ at 50 K to 0.120 \AA$^2$ at 300 K, 
i.e., by almost a factor of 2.
At given temperature and pressure, $(\Delta r)^2$ is larger for 
hydrogen than for oxygen,
but the difference between both decreases for increasing temperature.
Thus, at $T$ = 50 K and $P$ = 6 GPa, $(\Delta r)^2$ for hydrogen is
3.9 times the mean-square displacement of oxygen, whereas this
ratio decreases to 2.2 at 300 K.
This reflects the fact that the larger quantum delocalization of hydrogen
becomes comparatively less relevant as temperature increases.

The mean-square displacement $(\Delta r)^2$ of a given atomic nucleus
can be divided as $(\Delta r)^2 = Q_r^2 \, + \, C_r^2$, i.e., 
a ``purely quantum'' contribution $Q_r^2$ originating from the spread 
of the quantum paths [see Eq.~(\ref{qr2})], and another,
$C_r^2$, given by the spatial displacement of the centroid $\overline{\bf r}$
of the paths, that can be  derived from the motion of $\overline{\bf r}$
along a PIMD trajectory \cite{gi88,he11}:
\begin{equation}
 C_r^2 =  \left< \left( \overline{\bf r} - \langle \overline{\bf r} \rangle
                \right)^2 \right> 
       =  \langle  \overline{\bf r}^2 \rangle -
          \langle  \overline{\bf r} \rangle^2      \; .
\end{equation}
This part $C_r^2$ can be considered as a semiclassical thermal contribution
to $(\Delta r)^2$, as at high $T$ it converges to the mean-square
displacement given by a classical model.

At 50 K and 1 bar, $Q_r^2$ represents a 71\% of $(\Delta r)^2$ for hydrogen
and a 37\% for oxygen, since the quantum contribution
to the atomic delocalization is more important for hydrogen.
This can be also seen by comparing directly values of $Q_r^2$ for
both atomic species, which result to be 5.5 times larger for hydrogen
than for oxygen at $T$ = 50 K and $P$ = 1 bar.
At low temperatures, the quantum contribution $Q_r^2$ dominates the
spatial delocalization $(\Delta r)^2$, because $C_r^2$ converges to zero
as $T \to 0$ K.
The opposite happens at high temperatures, where the quantum contribution
$Q_r^2$ becomes small, eventually disappearing in the classical limit.
This high-$T$ limit, however, cannot be approached for ice VII due to the
onset of melting.

\begin{figure}
\vspace{-1.0cm}
\includegraphics[width= 8cm]{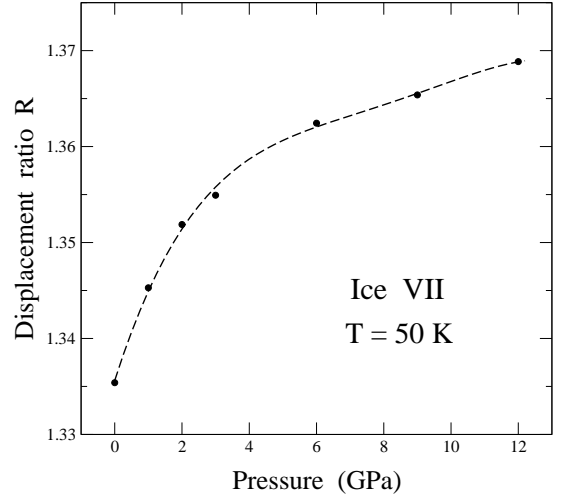}
\vspace{-0.3cm}
\caption{Ratio $R$ of mean-square displacements $(\Delta r)^2$ of
H and D in H$_2$O and D$_2$O ice VII, as a function of pressure at
$T$ = 50 K. The dashed line is a guide to the eye.
}
\label{f12}
\end{figure}

One can estimate a mean extension of the quantum paths of hydrogen from 
$Q_r^2$. At 50 K and 1 bar we find an average extension 
$\Delta x = \frac13 (Q_r^2)^{1/2} \approx$ 0.14 \AA,
which results to be much smaller than the H--H distance in a water molecule,
supporting the disregard of quantum exchange between protons in the PIMD 
simulations \cite{he11b,he14}.

In Fig.~11 we also present results for the mean-square displacement
$(\Delta r)^2$ of deuterium in D$_2$O ice VII. This displacement is smaller
than for H in H$_2$O ice VII, as expected for the larger mass of
deuterium.
In a purely harmonic approximation the mean-square displacement
of a nucleus at low $T$ (say $k_B T \ll \hbar \omega$, with $\omega$ a
typical frequency of the nuclear motion) scales with the mass $m$ as 
$(\Delta r)^2 \propto 1 / \sqrt{m}$.
In the high-temperature (classical) limit, $(\Delta r)^2$ is independent
of the mass.
Thus, when comparing $(\Delta r)^2$ of H and D in ice VII at 50 K,
we expect to find a ratio of mean-square displacements
$R = (\Delta r)^2_{\rm H} / (\Delta r)^2_{\rm D}$ close to $\sqrt{2} = 1.41$.
This is in fact what we found from the results of the PIMD simulations,
as shown in Fig.~12. In this plot we present the ratio $R$ of
mean-square displacements as a function of pressure at 50 K.
For $P$ = 1 bar, we found $R$ = 1.34. This ratio rises for
increasing pressure, and for $P$ = 12 GPa our simulations yielded
$R$ = 1.37, a value closer to $\sqrt{2}$. This increase in $R$ is in line 
with the observation that the description of vibrational motion with
a quasi-harmonic model becomes more accurate as pressure is raised
(see Sec.~III.A).

We note that the mean-square displacements of oxygen and hydrogen are
in fact not independent magnitudes, since both H and O participate in
several vibrational modes. The spacial displacement corresponding to
each atomic species in each mode depends on the mass of both kinds of
atoms. This means that $(\Delta r)^2$ for oxygen should be different
in H$_2$O and D$_2$O ice VII. This is what we found in the PIMD 
simulations, with $(\Delta r)^2_{\rm O}$ smaller in the second case.
Even though this difference is not very large, it is appreciable and
at 50 K it amounts to 5.5\% for $P$ = 1 bar and 3.5\% for 12 GPa.

\section{Summary}

In this paper we have presented results of PIMD simulations of ice VII in 
the isothermal-isobaric ensemble at several hydrostatic pressures.
For a given pressure, these quantum simulations have allowed us to study 
this solid water phase in the temperature range where it is stable or 
metastable. 
The results were compared with those derived from a QHA approximation,
as well as with those obtained from classical MD simulations.

The q-TIP4P/F potential model is known to describe fairly well various 
structural and thermodynamic properties of low-pressure ice phases. 
Here we have investigated 
its reliability to predict properties of high-pressure phases such 
as ice VII. We have found that this flexible interatomic potential
predicts rather well the molar volume of this ice phase at relatively low
pressures, but it yields volumes that deviate from those derived from 
diffraction experiments as pressure is increased. Thus, for 12 GPa the
molar volume is overestimated by about a 6\%.

The molar volume of ice VII decreases with increasing pressure, 
but the interatomic distances in the solid display a peculiar trend,
characteristic of condensed phases of water.
Thus, the distance between oxygen atoms in adjacent molecules decreases 
for increasing pressure, but the intramolecular O--H distance becomes larger 
as pressure is raised. 
This results from the intra-intermolecular geometric correlation present 
in ice, relating the covalent O--H bond length to the corresponding H-bond
geometry.  As a consequence, a larger molar volume causes a stronger 
intramolecular covalent bond.
This behavior of bond lengths as a function of pressure is similar to that
found when temperature is modified: covalent bonds become stronger and
H-bonds between water molecules weaken as $T$ rises.

The relevance of quantum effects has been assessed by comparing results
yielded by PIMD simulations with those given by classical MD simulations.  
Structural variables are found to change when quantum nuclear motion is
taken into account, especially at low temperatures. 
Thus, the molar volume and interatomic distances change appreciably 
in the range of pressure and temperature considered here.
At 1 bar and 50 K, the molar volume of ice VII is found to rise by
0.42 cm$^3$/mol (a 3.5\% of the classical value), and the intramolecular
O--H distance increases by 1.5\% due to quantum motion.

At a given temperature, the kinetic energy of hydrogen, $E_k$(H), given
by the q-TIP4P/F potential is found to increase for rising pressure.
However, the pressure dependence of $E_k$(H) changes critically with the
variation of the frequencies of stretching modes, which are known to
display negative Gr\"uneisen parameters. This potential model yields
a decrease of these $\omega_i$ for rising pressure, but it seems to
underestimate the value of the corresponding $\gamma_i$, which should
be more negative.

We have analyzed the influence of a hydrostatic pressure on
anharmonic effects in ice VII.
Our results indicate that the validity of the QHA to describe structural
and thermodynamic properties of this water phase increases as pressure is
raised. This is mainly a consequence of the relative importance of elastic
and vibrational energy, with the former becoming increasingly dominant
as pressure increases. Consequently, the precision requirements for
a good description of anharmonic vibrational modes is reduced with
rising pressure. 
It has been also argued that structural properties of solids under
pressure can be described rather accurately by a classical model for the
lattice vibrations.  The reason for this is similar to the decreasing
of anharmonicity as pressure rises, because in this respect
the actual description of the lattice vibrations by a classical
or a quantum model becomes less important for solids under high pressures.
This is questionable for spectroscopic features of high-pressure phases
when a certain accuracy is required, since vibrational frequencies 
predicted by a classical model or a QHA are not guaranteed to describe 
precisely the actual ones for high pressure.
All this applies to structural properties of an isolated ice phase, but
quantum motion of protons has been found to be relevant for high-pressure
transitions between ice phases, as in the H-bond symmetrization taking
place in the transition from ice VII to ice X.

We finally note that this kind of atomistic simulations of high-pressure
phases of water could be improved by the design of interatomic potentials
describing accurately their $P-V$ equation of state.
This may probably be achieved by adapting present-day potential models,
modifying the short-range intermolecular interactions to allow for a
larger compressibility at high pressures.

\begin{acknowledgments}
This work was supported by Direcci\'on General de Investigaci\'on,
MINECO (Spain) through Grant FIS2012-31713.
\end{acknowledgments}

\end{document}